\documentclass[preprint,showpacs,preprintnumbers,amsmath,amssymb,nofootinbib]{revtex4}
\usepackage{bbm}
\usepackage{amsfonts}
\usepackage{booktabs}
\usepackage{mathrsfs}
\usepackage{epsfig}
\usepackage{graphicx}
\usepackage{dcolumn}
\usepackage{bm}
\usepackage{amsmath}
\usepackage{slashed}
\usepackage{color}

\let\jnfont=\rm
\def\NPB#1,{{\jnfont Nucl.\ Phys.\ B }{\bf #1},}
\def\PLB#1,{{\jnfont Phys.\ Lett.\ B }{\bf #1},}
\def\EPJC#1,{{\jnfont Eur.\ Phys.\ Jour.\ C }{\bf #1},}
\def\PRD#1,{{\jnfont Phys.\ Rev.\ D }{\bf #1},}
\def\PRL#1,{{\jnfont Phys.\ Rev.\ Lett.\ }{\bf #1},}
\def\MPLA#1,{{\jnfont Mod.\ Phys.\ Lett.\ A }{\bf #1},}
\def\JPG#1,{{\jnfont J.\ Phys.\ G}{\bf #1},}
\def\CTP#1,{{\jnfont Commun.\ Theor.\ Phys.\ }{\bf #1},}
\def\ZPC#1,{{\jnfont Z.\ Phys.\ C }{\bf #1},}
\def\JHEP#1,{{\jnfont JHEP \ }{\bf #1},}
\def\Rv{\not{\hbox{\kern-1pt $R$}}}
\def\p{\not{\hbox{\kern-3pt $p$}}}

\begin{document}

\title{Revisiting Associated Production of 125 GeV Higgs Boson with a Photon at a Higgs Factory}

\author{Song Lin Hu$^{1}$, Ning Liu$^{1}$, Jie Ren$^{1}$, Lei Wu$^{2}$\\~ \vspace*{-0.3cm} }
\affiliation{ $^1$ Physics Department, Henan Normal University, Xinxiang 453007, China\\
$^2$ ARC Centre of Excellence for Particle Physics at the Terascale, School of Physics,
The University of Sydney, NSW 2006, Australia \vspace*{1.5cm}}

\begin{abstract}
Considering the constraints from the flavor physics, precision electroweak measurements, Higgs data and dark matter detections, we scan over the parameter space of the MSSM and calculate the cross section of $e^+e^- \to h \gamma$ in the allowed parameter space. Since the loop-induced gauge couplings $h\gamma\gamma$ and $hZ\gamma$ can simultaneously contribute to the process $e^+e^- \to h \gamma$, we find the cross section can be sizably enhanced by a light stau, maximally 1.47(1.38) times larger than the SM prediction at $\sqrt{s}=240(350)$ GeV. So with the high luminosity, the measurement of $e^+e^- \to h \gamma$ may be used to test the anomalous gauge couplings $h\gamma\gamma$ and $hZ\gamma$ in the MSSM at a Higgs Factory.
\end{abstract}
\maketitle

\section{Introduction}

The existence of a 125 GeV Higgs boson has been recently confirmed by the ATLAS and CMS collaborations at the LHC \cite{atlas,cms}. The next main step of the LHC searches is to discover new particles beyond the SM. As one of the most theoretically well-motivated scenarios for new physics, the Minimal Supersymmetric Standard Model(MSSM) have been widely studied by the theorists and experimentalists. However, up to now, the LHC has not found any evidences of the SUSY particles(sparticles). The negative results of direct searches for sparticles have pushed up the mass limits of the first two generation squarks and gluino into TeV region \cite{gluino}. The third generation squarks and non-colored sparticles have also been constrained in the simplified models \cite{exp-incl-ew,exp-st,exp-sb}. But they are still allowed to be at hundred GeV and may live in some hidden corners, due to their complicated decay modes \cite{reece,hidden,wu-stop,wu-higgsino}.

In contrast with the direct searches, an advantage of indirect searches lies in the fact that the results weakly depend on the kinematics configurations of sparticles. In this case, an alternative way is to find the indirect SUSY signals via loop corrections by the high precise measurements of the newly discovered Higgs boson at a Higgs factory. Among several proposals for the Higgs factories, the circular $e^+e^-$ colliders have been widely investigated, such as the TLEP \cite{hf1,hf2,tlep}. The proposed TLEP $e^+e^-$ collider \cite{tlep} could be located in a new 80 to 100 km tunnel in the Geneva area. It would be able to produce collisions at 4 interaction points with centre-of-mass energies from 90 to 350 GeV and beyond and is expected to make precision measurements at the $Z$ pole, at the $WW$ threshold, at the $HZ$ cross section maximum, and at the $t\bar{t}$ threshold, with an unprecedented accuracy. As comparison with other linear $e^+e^-$ collider, such as ILC and CLIC, the luminosity expected at TLEP is between a factor 5 and 3 orders of magnitude larger than that expected for a linear collider, at all centre-of-mass energies from the $Z$ pole to the $t\bar{t}$ threshold, where precision measurements are to be made, hence where the accumulated statistics will be a key feature. According to the expected high performance of the TLEP \cite{tlep}, about $O(10^6)$ Higgs bosons can be produced per year through the Higgs bremsstrahlung process at $\sqrt{s}\sim 240$ GeV with an integrated luminosity of 10000 fb$^{-1}$\cite{zh-loop1,zh-loop2,zh-loop3,hzgam}, which allows to measure the Higgs boson couplings at a percent level \cite{hf2,tao}. In addition to these studies, the TLEP also provides a unique opportunity to examine the various Higgs boson rare productions and decays.

In this paper, we investigate the associated production of the SM-like Higgs boson ($h$) with a hard photon in the MSSM at the TLEP under the current experimental constraints. Since the process $e^+ e^- \to h\gamma$ occurs at loop level, it will be sensitive to the contributions from the new particles. Such a process has been studied some time ago in Refs.\cite{barroso,replo,gabrielli,hollik} and recently studied in Ref.\cite{tcyuan} for (un)polarized electron and positron beams. Since the leading order of $e^+e^- \to h\gamma$ occurs at one-loop level induced by the electroweak coupling, the cross section is rather small, but the signal is very clean at a $e^+e^-$ collider. This allows for a reasonable hope to observe these events \cite{hrlep} if enough data is collected at a future high-energy collider. Given the expected high luminosity ${\cal L}=10,000$ fb$^{-1}$, there can be about 2,000 events to be obtained at the TLEP with $\sqrt{s}=240$ GeV. So we can expected that the TLEP may have a promising potential to detect the process $e^+e^- \to h\gamma$. But of course, the final feasibility study will depend on the future detector and analysis performance of the TLEP, which is beyond the scope of this work.

On the other hand, due to the recent constraints on the parameter space of the MSSM from the LHC experiments and the dark matter detections, it is necessary to reevaluate the size of the SUSY corrections to $e^+e^- \to h\gamma$ in the allowed parameter space. Besides, The process $e^+ e^- \to h\gamma$ can be used to probe the anomalous couplings of $h\gamma\gamma$ and $hZ\gamma$ \cite{carlos,wanglei,archil,huangda,hancheng}. At the LHC, most measurements of the properties of the Higgs boson are consistent with the SM expectations \cite{lhc-higgs}. However, the signal strength of diphoton decay mode reported by ATLAS is considerably larger than the SM prediction \cite{atlas-diphoton}, and this excess may persist in the future. In the MSSM, $h \to \gamma\gamma$ and $h \to Z \gamma$ can be simultaneously enhanced by a light stau with the large $\mu$ and $\tan\beta$ \cite{mssm-diphoton,mssm-zgam}, which will also lead to a significant enhancement in the process $e^+ e^- \to h\gamma$. Meanwhile, both ATLAS and CMS experiments have projected their sensitivities to high luminosities under various assumptions of detector and analysis performance in Ref.\cite{hf2}. For rare decays $h \to Z\gamma$, since the signal-background ratios is only about 0.5\%, the expected relative precisions on the signal strengths of $h \to Z\gamma$ at best reach about 20\% at 14 TeV LHC with ${\cal L} =3000$ fb$^{-1}$. However, according to the study of the Ref.\cite{hf2}, the production rate of $pp \to h \to Z\gamma$ in the MSSM at the LHC just is 1.1 times the SM prediction. So, it is very challenging for the HL-LHC to observe such indirect MSSM effects in the rare decay $h \to Z\gamma$. On the other hand, since there is no available studies of the expected ILC measurement of $h \to Z\gamma$, we can only estimate $h \to Z\gamma$ by referring to the ILC accuracy of $h \to \gamma \gamma$. At the ILC, the fast simulation studies indicate that $\sigma\cdot Br(h \to \gamma\gamma)$ can be measured with an accuracy of 34\% using $e^+e^- \to Zh$ at $\sqrt{s}=250$ GeV with a luminosity 250 fb$^{-1}$ \cite{ilc}. In the SM, the branching ratio of $h \to Z\gamma$ is about 1.5 times smaller than $h \to \gamma\gamma$ for $m_h=125$ GeV. This means the events number of $h \to Z\gamma$ collected by the ILC will be only about 2/3 of $h \to \gamma\gamma$. In addition, the multiplicity of final states in $h \to Z(\to f\bar{f})\gamma$ may reduce the reconstruction efficiency as comparison with $h \to \gamma\gamma$. So we can infer that the 250 GeV ILC accuracy for $h \to Z\gamma$ at most reach $34\%$ as the same as $h \to \gamma\gamma$, which is expected to be improved to 8.5\% for a luminosity 1000 fb$^{-1}$ at $\sqrt{s}=1000$ GeV \cite{ilc}. In this case, a super-high luminosity Higgs factory, such as TLEP, may be needed to detect the indirect MSSM effects in $h \to Z\gamma$.

The paper is organized as follows. In Sec. II. we briefly describe the scan of the parameter space of the MSSM and the calculations for the process $e^+e^- \to h\gamma$. In Sec.III we present the numerical results and discussions. Finally, we draw the conclusions in Sec. IV.

\section{scan methodology and calculation of $e^+e^- \to H\gamma$}
In this work, we will aim to examine the enhancement effects in $e^+e^- \to h\gamma$ and its correlation with the LHC signal strengths $R_{\gamma\gamma}$ and $R_{Z\gamma}$ in the MSSM. Given the previous results, we focus on the light stau scenario of MSSM, where the loop induced couplings $h\gamma\gamma$ and $hZ\gamma$ are found to be enhanced. Note that such a scenario is different from the natural MSSM with a light stop and light higgsinos. The latter one can be probed by searches for stop pair production \cite{wu-stop} or mono-jet signals induced by the degenerate higgsinos \cite{wu-higgsino}. While for our scenario, the direct searches for the light stau pair production or precise measurement of the rare Higgs decay $h \to \gamma\gamma$ can be used to test it at the LHC \cite{mssm-diphoton,mssm-zgam}. In the MSSM, after the electroweak symmetry breaking, there are two CP-even Higgs bosons($h,H$), one CP-odd Higgs boson($A$) and the charged Higgs bosons($H^{\pm}$). Although the mass of the lighter CP-even Higgs boson ($m_{h}$) is smaller than $M_{Z}$ at tree level, it can receive the large radiative corrections from the stop sector at one-loop level. The leading part of the corrections from the stop sector can be
expressed as \cite{mh-1loop}
\begin{eqnarray}
\Delta m^{2}_{h}(~\tilde{t}~)\simeq\frac{3m^{4}_{t}}{2\pi^{2}v^{2}\sin^{2}\beta}[\log
\frac{m_{\tilde{t}_{1}}m_{\tilde{t}_{2}}}{m^{2}_{t}}
+\frac{X^{2}_{t}}{2m_{\tilde{t}_{1}}m_{\tilde{t}_{2}}}(1-\frac{X^{2}_{t}}
{6m_{\tilde{t}_{1}}m_{\tilde{t}_{2}}})]
\end{eqnarray}
where $X_{t}\equiv A_{t}-\mu\cot\beta$ is the mixing parameter of stop. To increase $m_{h}$ to 125 GeV, it needs the heavy stop masses or a sizable stop mixing parameter $X_t$. In our study, we calculate the Higgs mass by using the package \textsf{FeynHiggs2.10.0} \cite{feynhiggs} and impose the collider constraints on the MSSM Higgs sector by using the package \textsf{HiggsBounds-4.1.0}\cite{higgsbounds}. In order to get the parameter space of the MSSM allowed by the current experiments, we scan the following parameter spcae:
\begin{eqnarray}
&&1 \le \tan\beta \le 60, ~~100 {~\rm GeV} \le M_A \le 1 {~\rm TeV}, ~~100 {~\rm GeV} \le \mu \le 2 {~\rm TeV}, \nonumber \\
&& 100 {~\rm GeV} \le \left(M_{Q_3},M_{U_3}\right) \le 2 {~\rm TeV},
~~ 100 {~\rm GeV} \le \left( M_{L_3},M_{E_3} \right)\le 1 {~\rm TeV}, \nonumber\\
&& -3 {~\rm TeV}\le A_t \le 3 {~\rm TeV},~~50 {~\rm GeV}\le M_{1} \le 500 {~\rm GeV} .
\label{MSSMscan}
\end{eqnarray}

We fix the first two generation squark soft masses($M_{\tilde{q}_{1,2}}$) and gluino mass($M_3$) at 2 TeV, and set $m_{U_3}=m_{D_3}$, $A_t = A_b$. We take the grand unification relation $3 M_1/5 \alpha_1 = M_2/\alpha_2$ for electroweak gaugino masses. Since the first two generation sleptons are irrelevant to our study, we decouple them for simplicity. According to Ref.\cite{mssm-zgam}, we take $A_{\tau}=0$ in our calculations. The reason is that the enhancement of $h \to \gamma\gamma, Z\gamma$ is sensitive to the stau mass and $\mu\times\tan\beta$. So without large $A_{\tau}$, one can also have a light stau by setting the relevant soft mass $M_{L_3}$ and $M_{E_3}$.

In the scan, we consider the following experimental and theoretical constraints:
\begin{enumerate}
   \item We require that the mass of the light CP-even Higgs for each samples be in the region of 123 GeV$<m_{h}<$ 127 GeV \cite{feynhiggs}. At the same time, the current bounds for the Higgs bosons from LEP, Tevatron and LHC should be satisfied \cite{higgsbounds};
   \item  We require our samples to satisfy the B-physics bounds at 2$\sigma$ level, including $B\rightarrow X_s\gamma$ and the latest measurements of $B_s\rightarrow \mu^+\mu^-$, $B_d\rightarrow X_s\mu^+\mu^-$ and $ B^+\rightarrow \tau^+\nu$. We use the package of \textsf{SuperIso v3.3} \cite{superiso} to implement these constraints;
   \item By using the package of \textsf{MicrOMEGAs v2.4} \cite{micromega}, we impose the dark matter constraints of the neutralino relic density from PLANCK(in 2$\sigma$ range) \cite{planck} and the direct detection results from LUX (at 90\% confidence level) \cite{lux};
   \item Since the large mixing terms in the stau sector will jeopardize the vacuum stability in the MSSM \cite{vev0,vev1,vev2,vev3}, our samples are required to comply with the vacuum meta-stability condition by using the formula in Ref. \cite{vev3}:
       \begin{eqnarray}
|\mu \tan \beta_{\textrm{eff}}| < 56.9 \sqrt{m_{\tilde{\tau}_L} m_{\tilde{\tau}_R}} + 57.1 \left(m_{\tilde{\tau}_L}+1.03 m_{\tilde{\tau}_R} \right)   - 1.28 \times 10^4 ~{\rm GeV} \nonumber \\
+\frac{1.67 \times 10^6 ~{\rm GeV} ^2 }{m_{\tilde{\tau}_L}+m_{\tilde{\tau}_R} }  - 6.41 \times 10^7 ~{\rm GeV} ^3 \left ( \frac{1}{m_{\tilde{\tau}_L}^2  } + \frac{0.983}{m_{\tilde{\tau}_R}^2}  \right) . \label{metabound}
\end{eqnarray}

 \end{enumerate}

We mainly use the flat scan method to obtain the samples that satisfy our constraints. But according to the previous studies \cite{mssm-diphoton,mssm-zgam}, we extra generate more random points in the region with large $\mu\times\tan\beta$ and light stau to obtain the loop induced coupling $h\gamma\gamma$ and $hZ\gamma$ as large as possible. Such method can successfully cover most of the parameter space that relevant for our study, which has also been used in our group work \cite{mssm-zgam} and been cross checked by other group \cite{nmssm-zgam}. After our scan, we obtain 1078 samples allowed by the listed constraints (1)-(4) in the above. The values of $m_A$ for these samples are larger than about $350$ GeV. Since there is still a large discrepancy between the SM prediction and experimental results \cite{g-2}, we do not require our samples to explain the anomalous muon $g-2$ in our calculations. But it should be mentioned that the large $\mu\times\tan\beta$ will be helpful for alleviating this tension in the MSSM.

In the MSSM, the process $e^+e^- \to h \gamma$ includes the following subprocesses: $(i)$ $s$--channel: $\gamma,Z$ vertex diagrams that are corrected by the charged Higgs boson, chargino, squark and slepton; $(ii)$ $t$--channel: $hee$ vertex diagrams that are corrected by chargino/sneutrino and neutralino/selectron; $(iii)$ box diagrams that involve neutralino/selectron and chargino/sneutrino states. We denote the four-momenta of initial and final states in the process as
\begin{eqnarray}
e^+(q_1) + e^-(q_2) \rightarrow h(p_3)+ \gamma(p_4)\label{eehr}
\end{eqnarray}
All the amplitudes of Eq.(\ref{eehr}) are generated by \textsf{FeynArts-3.9} \cite{feynart}, and are further reduced by \textsf{FormCalc-8.3} \cite{formcalc}. The numerical calculations are performed by using \textsf{LoopTools-2.8} \cite{looptools}. In order to preserve supersymmetry, we adopt the constrained differential renormalization (CDR) \cite{cdr1} to regulate the ultraviolet divergence (UV) in the virtual corrections, which is equivalent to the dimensional reduction method at one-loop level \cite{cdr2}. We numerically checked the UV cancellation and notice that the $Z-\gamma$ self-energy mixing term is required to get the finite results. Note that the infrared singularities may occur in the $t-$channel $hee$ vertex and $W/Z$ box diagrams. However, since we keep the electron mass in the loop functions, infrared singularities that appear as logarithmic singularities $ln(m_e)$ can cancel exactly when all these diagrams are summed. We also checked our results with those of Ref. \cite{hollik} by setting the same SM parameters and found they are consistent well. In order to show the SUSY effects in the process $e^+ e^- \to h \gamma$, we define the following ratio:
\begin{eqnarray}
R_{h\gamma} & \equiv & \frac{\sigma_{MSSM}(e^+e^-\to h\gamma)}{\sigma_{SM}(e^+e^-\to h\gamma)}.
\end{eqnarray}

\section{Numerical Result and discussions}
\begin{figure}[ht]
\centering
\includegraphics[width=6cm,height=5cm]{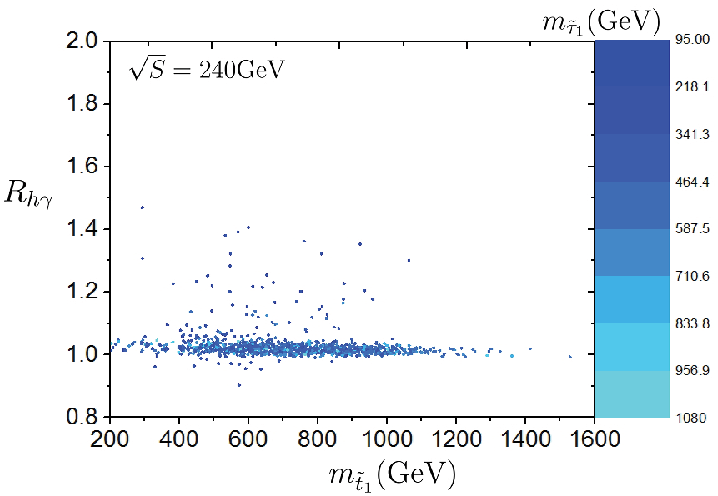}
\includegraphics[width=6cm,height=5cm]{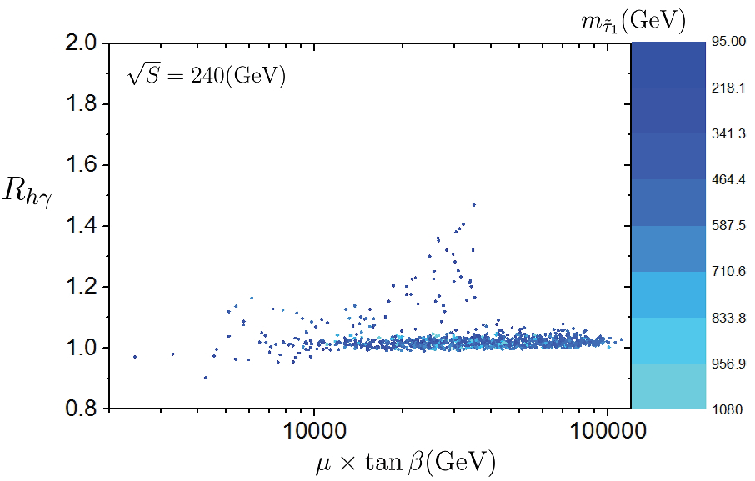}
\includegraphics[width=6cm,height=5cm]{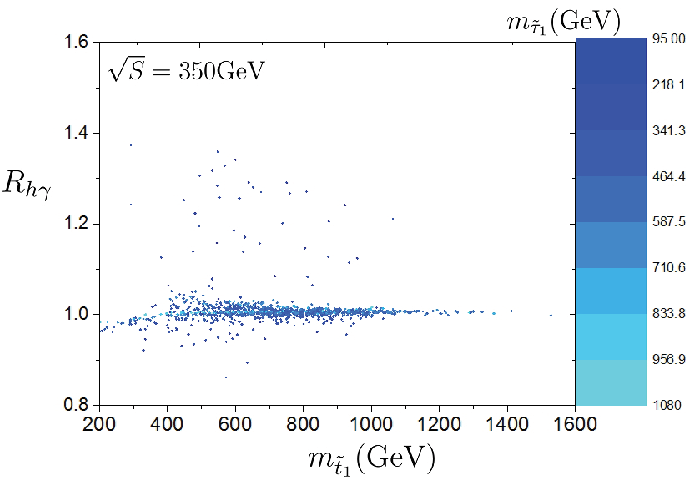}
\includegraphics[width=6cm,height=5cm]{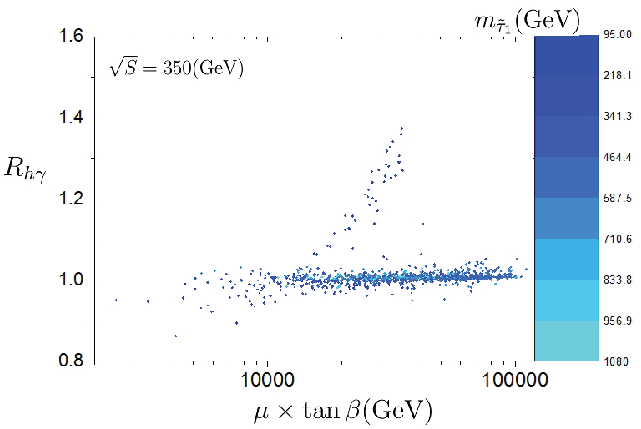}
\vspace*{-0.5cm}
\caption{The dependence of the ratio $R_{h\gamma}$ on the lighter stop mass($m_{\tilde{t}_1}$), the lighter stau mass($m_{\tilde{\tau}_1}$) and $\mu\tan\beta$ at the TLEP with $\sqrt s=240,350$ GeV.}\label{fig1}
\end{figure}
In our numerical calculations, we take the input parameters of the SM as \cite{pdg}
\begin{eqnarray}
m_t=173.5{\rm ~GeV}, ~~m_{e}=0.519991{\rm ~MeV}, ~~m_{Z}=91.19 {\rm
~GeV},\nonumber
\\~~\sin^{2}\theta_W=0.2228, ~~\alpha(m_Z^2)^{-1}=127.918.~~~~~~~~~~~~~~~~
\end{eqnarray}
We keep the same Higgs mass for the calculation of $e^+e^- \to h\gamma$ in the SM and MSSM. In addition, since we use the loop-corrected Higgs mass in the phase space, we should also correct all involved Higgs couplings in order to get the SM result of $e^+e^- \to h\gamma$ in the decoupling limit. Such calculations can be done by correcting the mixing angle $\alpha$ and the CP-even Higgs masses \cite{hollik}. It should be noted that the only one exception is the trilinear couplings of the CP-even Higgs boson couplings to charged Higgs bosons and Goldstones that could not be completely mapped into the corrections to the angle $\alpha$. We correct these couplings as the Ref.\cite{correction}



In Fig.\ref{fig1}, we present the dependence of the ratio $R_{h\gamma}$ on the lighter stop mass($m_{\tilde{t}_1}$), the lighter stau mass($m_{\tilde{\tau}_1}$) and $\mu\times\tan\beta$ at the TLEP with $\sqrt{s}=240,350$ GeV. We can see that the large values of $R_{h\gamma}$ are obtained when the masses of sparticles involving in the loop become small. Due to the $s$-channel suppression, the cross section of $e^+e^- \to h\gamma$ for $\sqrt{s}=350$ GeV is smaller than the one for $\sqrt{s}=240$ GeV. The maximal value of $R_{h\gamma}$ can be 1.47(1.38) times larger than the SM prediction at $\sqrt{s}=240(350)$ GeV, which corresponds to the cross section to be 0.147(0.0493) fb in the MSSM. When the stop mass becomes heavy, the value of $R_{h\gamma}$ will be small but can still be enhanced by a light stau with large $\mu\times\tan\beta$. The reason is that the dominant contribution of sfermions to $e^+ e^- \to h\gamma$ comes from the stau loop, which can be understood from the followings: the leading part of the amplitudes of the sfermions loop is proportional to $(g_a A_{\tilde{f}}+ g_b \mu\times\tan\beta) \sin 2 \theta_{\tilde{f}}/m_{\tilde{f}_1}^2$ \cite{hollik}. To satisfy the requirement of the Higgs mass, heavy stops or a large mixing parameter $A_t$ is needed. The light $\tilde{t}_1$ can be obtained by the large $A_t$ but accompanies with a heavy $\tilde{t}_2$. This will lead to a small stop mixing angle $\theta_{\tilde{t}}$ and reduce the stop loop contribution. However, the light stau can be achieved without a large $A_\tau$ by setting the relevant soft mass parameters $M_{L_3}$ and $M_{E_3}$ \cite{mssm-diphoton}.  So only a light stau with a large value of $\mu\times\tan\beta$ can sizably contribute to the process $e^+ e^- \to h\gamma$.

\begin{figure}[ht]
\centering
\includegraphics[width=7cm,height=6cm]{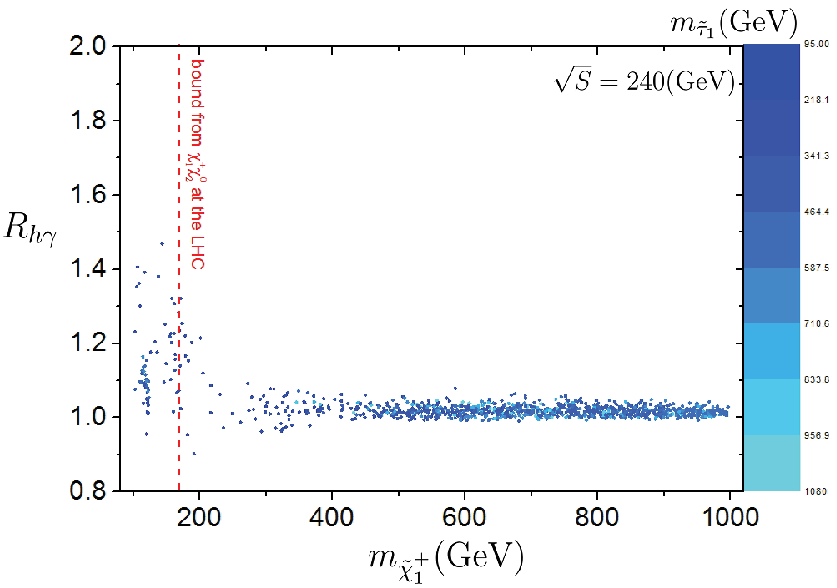}
\includegraphics[width=7cm,height=6cm]{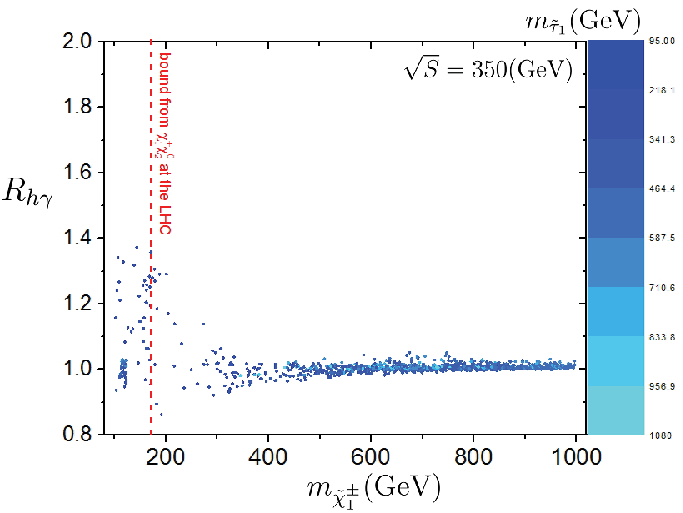}
\vspace*{-0.5cm}
\caption{The dependence of the ratio $R_{h\gamma}$ on the lightest chargino mass($\tilde{\chi}^{\pm}_{1}$) and the lighter stau mass($m_{\tilde{\tau}_1}$) at the TLEP with $\sqrt s=240,350$ GeV. The bound on the $m_{\tilde{\chi}^{+}_{1}}$ is taken from the right panel of Fig.8 in Ref. \cite{3l}.}\label{fig2}
\end{figure}
In Fig.\ref{fig2}, we show the effects of $m_{\tilde{\chi}^{\pm}_{1}}$ and $m_{\tilde{\tau}_1}$ in the ratio $R_{h\gamma}$. We can see that when the mass of $\tilde{\chi}^+_1$ is below 400 GeV, the mass of $\tilde{\tau}_1$ is always smaller than about 200 GeV. This is because that such a light stau is needed to co-annihilate with the neutralino dark matter($\tilde{\chi}^0_1$) to guarantee the correct dark matter relic density \cite{light-stau-1,light-stau-2}. Due to the imposed GUT relation between $M_1$ and $M_2$, the dominant component of LSP in our case is bino-like with a mass larger about 50 GeV. The LUX exclusion can probe deeply into the Higgsino and Wino components. After imposing LUX, the wino and higgsino fractions of the LSP for our surviving samples are further constrained and each of them consists of less than 7\%. We also find that the contribution of light $\tilde{\chi}^+_1$ loop is much smaller than the light stau loop. It is because that the coupling $C_{h\tilde{\chi}^+_1\tilde{\chi}^-_1}$ is determined by the component of $\tilde{\chi}^+_1$, which can be large only when $\tilde{\chi}^+_1$ is a mixture of higgsino and wino \cite{susyint}. But in the mass range $m_{\tilde{\chi}^+_1} \lesssim 400$ GeV, we find that $\tilde{\chi}^+_1$ is dominated by wino, which will highly suppress the contribution of $\tilde{\chi}^+_1$ loop. Besides, the light wino-like chargino in our study may be constrained by the latest results from the searches for $\tilde\chi^{\pm}_{1}\tilde\chi^{0}_{2}$ at the LHC \cite{3l}. In our calculations, since we decouple the contributions of the first two generations of slepton and assume the GUT relation between $M_1$ and $M_2$, we estimate the impact of this bound on $R_{h\gamma}$ by simply using the result for $m_{\tilde{\chi}^{2}_{0}}=2m_{\tilde{\chi}^{1}_{0}}$ on the right panel of Fig.8 in the ATLAS paper \cite{3l}, where $\tilde{\chi}^{0}_{2}$ and $\tilde{\chi}^{\pm}_{1}$ are assumed to be pure winos. We can see that the values of $R_{h\gamma}$ are slightly reduced by this constraint. However, it should be noted that this direct search bound can become weak when a small portion of higgsino is involved in $\tilde{\chi}^{0}_{2}$ and $\tilde{\chi}^{\pm}_{1}$. So, given these considerations, we do not impose this constraint in our study.

\begin{figure}[ht]
\centering
\includegraphics[width=8cm,height=7cm]{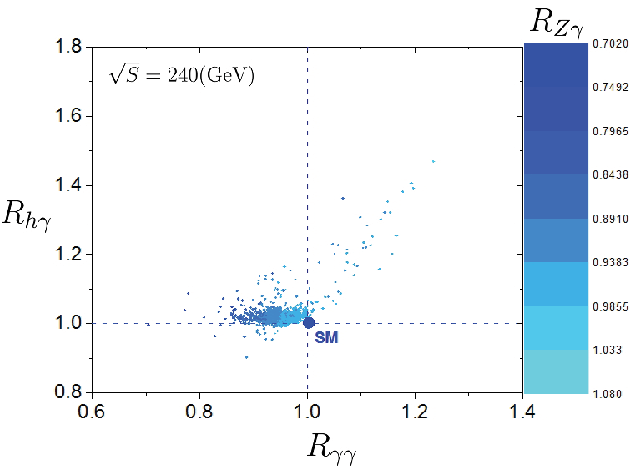}
\includegraphics[width=8cm,height=7cm]{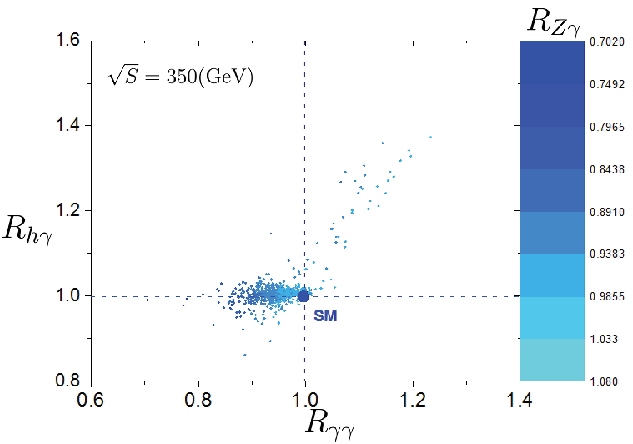}
\vspace*{-0.5cm}
\caption{The correlations of $R_{h\gamma}$ at the TLEP for $\sqrt{s}=240,350$ GeV with $R_{\gamma\gamma}$ and $R_{Z\gamma}$ at the LHC.}\label{fig3}
\end{figure}
In Fig.\ref{fig3}, we study the correlation of $R_{h\gamma}$ at the TLEP with the signal strengthes $R_{\gamma\gamma}$ and $R_{Z\gamma}$ at the LHC, which are defined as followings:
\begin{eqnarray}
R_{\gamma\gamma} & \equiv &  \frac{\sigma_{MSSM}(pp\to h\to \gamma\gamma)}{\sigma_{\rm SM}(pp\to h\to \gamma\gamma)},\\
R_{Z\gamma} & \equiv &  \frac{\sigma_{MSSM}(pp\to h\to Z\gamma)}{\sigma_{\rm SM}(pp\to h\to Z\gamma)}.
\end{eqnarray}
We adopt the narrow width approximation method to calculate the processes $pp \to h \to \gamma\gamma, Z\gamma$. Since the dominant production is from the gluon fusion, the total cross section $\sigma(pp \to h \to \gamma\gamma,Z\gamma)$ can be approximately written as $\sigma(pp \to h \to \gamma\gamma,Z\gamma)=\sigma(gg \to h)\times Br(h \to \gamma\gamma,Z\gamma)$. In the calculation of $gg \to h$, we use 2-loop evolution for the strong coupling constant $\alpha_s(\mu)$ with QCD parameter $\Lambda^{n_{f}=5}=226{\rm ~MeV}$ and obtain $\alpha_s(m_Z)=0.118$. We take CTEQ6M as the parton distribution functions (PDF) \cite{cteq} and set the renormalization scale $\mu_R$ and factorization scale $\mu_F$ to be $\mu_R=\mu_F=m_h$. Note that, the dependence of results on the PDF and scale setting is very weak because of the cancelation between numerator and denominator in the ratios. It can be seen that the ratio $R_{h\gamma}$ is approximately proportional to $R_{\gamma\gamma}$ and $R_{Z\gamma}$. The reason is that the loop-induced couplings $h\gamma\gamma$ and $hZ\gamma$ can simultaneously contribute to the process $e^+e^- \to h\gamma$, when both of $R_{\gamma\gamma}$ and $R_{Z\gamma}$ get the enhancement from light stau, $R_{h\gamma}$ can be significantly enhanced and maximally reach about 1.47(1.38) at the TLEP with $\sqrt{s}=240(350)$ GeV. This correlation behavior is different from some non-SUSY models predictions \cite{huangda,hancheng}, and may be used as complementary way to discriminate new physics models. Therefore, the measurement of $e^+e^- \to h\gamma$ at the TLEP may be helpful to test the anomalous gauge couplings $h\gamma\gamma$ and $hZ\gamma$ in the MSSM.

\section{SUMMARY AND CONCLUSION}
In spite of the discovery of the Higgs boson, the precise measurement of the Higgs boson and the searches for new physics beyond the SM are just starting. In particular, the rare productions and decays of the Higgs boson that sensitive to the new physics are worth thoroughly exploring at the future colliders. However, given the limited capability of the LHC, it is necessary to consider those rare processes at a $e^+e^-$ collider. In this paper, we investigate the rare production process $e^+e^- \to h\gamma$ that involves two sensitive loop induced couplings $h\gamma\gamma$ and $hZ\gamma$ in the MSSM at a future Higgs factory, such as TLEP with $\sqrt{s}=240,350$ GeV. We found that the MSSM corrections can enhance the cross section to be 0.147(0.0493) fb by a light stau with large $\tan\beta\times \mu$ at $\sqrt{s}=240(350)$ GeV. By virtue of the high luminosity of TLEP, such a rare process may have a potential to be observed at TLEP. Besides, we analyzed the correlation between the $e^+e^-$ collider signal strength $R_{h\gamma}$ and the LHC signal strengths $R_{\gamma\gamma}$ and $R_{Z\gamma}$. We found that they have a strong positive correlation, which is different from some non-SUSY models predictions. So such a correlation behavior may play a complementary role to discriminate new physics models in the future.

\section*{Acknowledgement}
We appreciate Dr. Teppei Kitahara for introducing the vacuum stability constraints on the light stau sector in the MSSM. We also thank the discussions with Jin Min Yang, Junjie Cao, Chengcheng Han and Peiwen Wu. Ning Liu would like to thank Dr. Archil Kobakhidze for his warm hospitality in Sydney node of CoEPP in Australia. This work is supported by the Australian Research Council, the National Natural Science Foundation of China (NNSFC) under grant Nos.11305049, 11275057, 11275245, 10821504 and 11135003, by Specialized Research Fund for the Doctoral Program of Higher Education under Grant No.20134104120002, and by the Startup Foundation for Doctors of Henan Normal University under contract No.11112.


\begin{thebibliography}{90}

\bibitem{atlas}
  G.~Aad {\it et al.}  [ATLAS Collaboration],
  Phys.\ Lett.\ B {\bf 716}, 1 (2012).

\bibitem{cms}
  S.~Chatrchyan {\it et al.}  [CMS Collaboration],
  Phys.\ Lett.\ B {\bf 716}, 30 (2012).


\bibitem{gluino}
ATLAS Collaboration, ATLAS-CONF-2013-054; ATLAS-CONF-2013-061;
CMS Collaboration, SUS-CONF-2013-007; SUS-CONF-2013-004.

\bibitem{exp-incl-ew}
ATLAS Collaboration, ATLAS-CONF-2013-047;
ATLAS-CONF-2013-049;
ATLAS-CONF-2013-036;
CMS Collaboration, CMS-PAS-SUS-13-006;
CMS-PAS-SUS-12-016;
CMS-PAS-SUS-12-027.

\bibitem{exp-st}
ATLAS Collaboration, ATLAS-CONF-2013-037;
ATLAS-CONF-2013-024.

\bibitem{exp-sb} ATLAS Collaboration, ATLAS-CONF-2013-053.

\bibitem{reece}
  J.~Fan, M.~Reece and J.~T.~Ruderman,
  JHEP {\bf 1111}, 012 (2011)  [arXiv:1105.5135 [hep-ph]].

\bibitem{hidden}
  M.~J.~Strassler,
  hep-ph/0607160;
  M.~Luo and S.~Zheng,
  JHEP {\bf 0904}, 122 (2009)  [arXiv:0901.2613 [hep-ph]].


\bibitem{wu-stop}
  J.~Cao, {\it et al.},
  JHEP {\bf 1211}, 039 (2012);
  C.~Han, K.~-i.~Hikasa, L.~Wu, J.~M.~Yang and Y.~Zhang,
  JHEP {\bf 1310}, 216 (2013)  [arXiv:1308.5307 [hep-ph]].

\bibitem{wu-higgsino}
  C.~Han, A.~Kobakhidze, N.~Liu, A.~Saavedra, L.~Wu and J.~M.~Yang,
  arXiv:1310.4274 [hep-ph];


\bibitem{hf1}
  A.~Blondel {\it et al.},
  arXiv:1302.3318 [physics.acc-ph].

\bibitem{hf2}
  S.~Dawson {\it et al.},
  arXiv:1310.8361 [hep-ex].

\bibitem{tlep}
  M.~Koratzinos, A.~P.~Blondel, R.~Aleksan, O.~Brunner, A.~Butterworth, P.~Janot, E.~Jensen and J.~Osborne {\it et al.},
  arXiv:1305.6498 [physics.acc-ph];
  M.~Bicer {\it et al.}  [TLEP Design Study Working Group Collaboration],
  JHEP {\bf 1401}, 164 (2014)
  [arXiv:1308.6176 [hep-ex]].

\bibitem{zh-loop1}
  J.~Fleischer and F.~Jegerlehner,
  Nucl.\ Phys.\ B {\bf 216}, 469 (1983);

\bibitem{zh-loop2}
  B.~A.~Kniehl,
  Z.\ Phys.\ C {\bf 55}, 605 (1992).

\bibitem{zh-loop3}
A.~Denner, J.~K\"ublbeck, R.~Mertig and M.~B\"ohm,
Z.\ Phys.\ C {\bf 56} (1992) 261.


\bibitem{hzgam}
  N.~Liu, J.~Ren, L.~Wu, P.~Wu and J.~M.~Yang,
  arXiv:1311.6971 [hep-ph].


\bibitem{tao}
  S.~B.~Giddings, T.~Liu, I.~Low and E.~Mintun,
  Phys.\ Rev.\ D {\bf 88}, 095003 (2013)  [arXiv:1301.2324 [hep-ph]].

\bibitem{barroso}
A. Barroso, J. Pulido, J. C. Romao, Nucl.\ Phys.\ B {\bf 267}, 509 (1985).

\bibitem{replo}
A. Abbasabadi, D. Bowser-Chao, D. A. Dicus and W. A. Repko, Phys.\ Rev.\ D {\bf 52}, 3919 (1995).

\bibitem{hollik}
  A.~Djouadi, V.~Driesen, W.~Hollik and J.~Rosiek,
  Nucl.\ Phys.\ B {\bf 491}, 68 (1997)  [hep-ph/9609420].


\bibitem{gabrielli}
  E.~Gabrielli, V.~A.~Ilyin and B.~Mele,
  hep-ph/9707370.

\bibitem{tcyuan}
  A.~Arhrib, R.~Benbrik and T.~-C.~Yuan,
  arXiv:1401.6698 [hep-ph].


\bibitem{hrlep}
P. M\"{a}ttig, Report CERN/95-081.

\bibitem{carlos}
  A.~Joglekar, P.~Schwaller and C.~E.~M.~Wagner,
  JHEP {\bf 1212}, 064 (2012)  [arXiv:1207.4235 [hep-ph]];

\bibitem{wanglei}
  L.~Wang and X.~-F.~Han,
  arXiv:1303.4490 [hep-ph].

\bibitem{archil}
  A.~Kobakhidze,
  arXiv:1208.5180 [hep-ph].


\bibitem{huangda}
  C.~-S.~Chen, C.~-Q.~Geng, D.~Huang and L.~-H.~Tsai,
  hys.\ Rev.\ D {\bf 87}, 075019 (2013)  [arXiv:1301.4694 [hep-ph]].
  C.~-W.~Chiang and K.~Yagyu,
  Phys.\ Rev.\ D {\bf 87}, no. 3, 033003 (2013)  [arXiv:1207.1065 [hep-ph]].

\bibitem{hancheng}
  C.~Han, N.~Liu, L.~Wu, J.~M.~Yang and Y.~Zhang,
  arXiv:1212.6728.




\bibitem{lhc-higgs}
The ATLAS Collaboration,
  ATLAS-CONF-2013-034.
The CMS Collaboration,
  CMS-PAS-HIG-13-005.

\bibitem{atlas-diphoton}
  ATLAS-CONF-2013-012.

\bibitem{mssm-diphoton}
  M.~Carena, S.~Gori, N.~R.~Shah, C.~E.~M.~Wagner and L.~-T.~Wang,
  JHEP {\bf 1207}, 175 (2012)  [arXiv:1205.5842 [hep-ph]].
  J.~Cao, Z.~Heng, T.~Liu and J.~M.~Yang,
  Phys.\ Lett.\ B {\bf 703}, 462 (2011)  [arXiv:1103.0631 [hep-ph]];
  A.~Gutierrez-Rodriguez, J.~Montano and M.~A.~Perez,
  J.\ Phys.\ G {\bf 38}, 095003 (2011)  [arXiv:1009.4354 [hep-ph]];
  J.~-J.~Cao, Z.~-X.~Heng, J.~M.~Yang, Y.~-M.~Zhang and J.~-Y.~Zhu,
  JHEP {\bf 1203}, 086 (2012)  [arXiv:1202.5821 [hep-ph]];
  J.~Cao, Z.~Heng, J.~M.~Yang and J.~Zhu,
  JHEP {\bf 1210}, 079 (2012)  [arXiv:1207.3698 [hep-ph]];
  R.~Sato, K.~Tobioka and N.~Yokozaki,
  Phys.\ Lett.\ B {\bf 716}, 441 (2012)
  [arXiv:1208.2630 [hep-ph]];


\bibitem{mssm-zgam}
  J.~Cao, L.~Wu, P.~Wu and J.~M.~Yang,
  JHEP {\bf 1309}, 043 (2013)  [arXiv:1301.4641 [hep-ph]];

\bibitem{ilc}
  D.~M.~Asner, T.~Barklow, C.~Calancha, K.~Fujii, N.~Graf, H.~E.~Haber, A.~Ishikawa and S.~Kanemura {\it et al.},
  arXiv:1310.0763 [hep-ph].

\bibitem{mh-1loop}
Y.~Okada, M.~Yamaguchi and T.~Yanagida, Prog.\ Theor.\ Phys.\  85, 1 (1991);
J.R.~Ellis, G.~Ridolfi and F.~Zwirner, Phys.\ Lett.\ B257, 83 (1991);
H.E.~Haber and R.~Hempfling, Phys.\ Rev.\ Lett.\ 66,  1815 (1991).

\bibitem{feynhiggs}
  M.~Frank,  {\it et al.},
  JHEP {\bf 0702}, 047 (2007);
  G.~Degrassi,  {\it et al.},
  Eur.\ Phys.\ J.\ C {\bf 28}, 133 (2003);
  S.~Heinemeyer, W.~Hollik and G.~Weiglein,
  Comput.\ Phys.\ Commun.\  {\bf 124}, 76 (2000);
  Eur.\ Phys.\ J.\ C {\bf 9}, 343 (1999);

\bibitem{higgsbounds}
  P.~Bechtle,  {\it et al.},
  Comput.\ Phys.\ Commun.\  {\bf 182}, 2605 (2011);
  Comput.\ Phys.\ Commun.\  {\bf 181}, 138 (2010).

\bibitem{superiso}
  F.~Mahmoudi,
  Comput.\ Phys.\ Commun.\  {\bf 180}, 1579 (2009);
  Comput.\ Phys.\ Commun.\  {\bf 178}, 745 (2008).


\bibitem{micromega}
  G.~Belanger, {\it et al.},
  Comput.\ Phys.\ Commun.\  {\bf 182}, 842 (2011).

\bibitem{planck}
  P.~A.~R.~Ade {\it et al.}  [Planck Collaboration],
  arXiv:1303.5076 [astro-ph.CO].

\bibitem{lux}
  D.~S.~Akerib {\it et al.}  [LUX Collaboration],
  arXiv:1310.8214 [astro-ph.CO].


\bibitem{vev0}
 J.~Hisano and S.~Sugiyama,
  Phys.\ Lett.\ B {\bf 696}, 92 (2011)  [Erratum-ibid.\ B {\bf 719}, 472 (2013)]  [arXiv:1011.0260 [hep-ph]].


\bibitem{vev1}
  M.~Carena, S.~Gori, I.~Low, N.~R.~Shah and C.~E.~M.~Wagner,
  JHEP {\bf 1302}, 114 (2013)  [arXiv:1211.6136 [hep-ph]].

\bibitem{vev2}
  T.~Kitahara,
  JHEP {\bf 1211}, 021 (2012)
  [arXiv:1208.4792 [hep-ph]].

\bibitem{vev3}
  T.~Kitahara and T.~Yoshinaga,
  arXiv:1303.0461 [hep-ph].

\bibitem{nmssm-zgam}
  G.~Belanger, V.~Bizouard and G.~Chalons,
  Phys.\ Rev.\ D {\bf 89}, 095023 (2014)
  [arXiv:1402.3522 [hep-ph]].

\bibitem{g-2}
 M.~Davier, A.~Hoecker, B.~Malaescu, C.~Z.~Yuan and Z.~Zhang,
  Eur.\ Phys.\ J.\ C {\bf 66}, 1 (2010)
  [arXiv:0908.4300 [hep-ph]].



\bibitem{feynart}
T. Hahn, Comput. Phys. Commun. 140, 418 (2001).

\bibitem{formcalc}
T. Hahn, M. Perez-Victoria, Comput. Phys. Commun. 118, 153 (1999).

\bibitem{looptools}
G. J. van Oldenborgh, Phys Commun 66, 1 (1991).

\bibitem{cdr1}
F.~del Aguila, A.~Culatti, R.~Tapia, and M.~Perez-Victoria,
Nucl. Phys. B537, 561 (1999).

\bibitem{cdr2}
W.~Siegel, Phys. Lett. {\bf B84}, 193 (1979); T.~Hahn and
M.~Perez-Victoria, Comput. Phys. Commun. {\bf 118}, 153 (1999),
hep-ph/9807565.


\bibitem{pdg} J. Beringer {\it et al.}, Particle Data Group, Phys.\ Rev. \ D {\bf 86}, 010001 (2012).


\bibitem{correction}
V. Barger, M.S. Berger, A.L. Stange and R.J.N. Phillips, Phys. Rev. D45 (1992)
4128; A. Brignole and F. Zwirner, Phys. Lett. B299 (1993) 72.

\bibitem{light-stau-1}
  J.~R.~Ellis, T.~Falk and K.~A.~Olive,
  Phys.\ Lett.\ B {\bf 444}, 367 (1998)  [hep-ph/9810360].

\bibitem{light-stau-2}
  J.~R.~Ellis, T.~Falk, K.~A.~Olive and M.~Srednicki,
  Astropart.\ Phys.\  {\bf 13}, 181 (2000)  [Erratum-ibid.\  {\bf 15}, 413 (2001)]  [hep-ph/9905481].

\bibitem{susyint}
J.~F.~Gunion and H.~E.~Harber, \NPB272, 1 (1986).

\bibitem{3l}
ATLAS Collaboration [ATLAS Collaboration], ATLAS-CONF-2013-035.

\bibitem{cteq}
P.~M.~Nadolsky, H.~-L.~Lai, Q.~-H.~Cao, J.~Huston, J.~Pumplin,
D.~Stump, W.~-K.~Tung and C.~-P.~Yuan,
Phys.\ Rev.\ D {\bf 78}, 013004 (2008), arXiv:0802.0007 [hep-ph].

\end{thebibliography}
\end{document}